\documentclass{article}
\usepackage{epsfig}
\newtheorem{guess}{Theorem}
\newtheorem{defn}{Definition}

\newcommand{\braX}{\langle x|}
\newcommand{\ketX}{|x\rangle}

\newcommand{\ketzero}{|0\rangle}
\newcommand{\ketone}{|1\rangle}

\title{Finding Solutions to NP Problems: Philosophical Differences 
Between Quantum and Evolutionary Search Algorithms}

\author{G. W. Greenwood}
\date{}

\begin{document}

\maketitle

\bibliographystyle{unsrt}

\begin{abstract}
There is no known polynomial-time algorithm that can solve an
NP problem.  Evolutionary search has been shown to be a viable
method of finding acceptable solutions within a reasonable
time period.  Recently quantum computers have surfaced as
another alternative method.  But these two methods use
radically different philosophies for solving NP problems even
though both search methods are non-deterministic.  This paper
uses instances of {\bf SAT}, {\bf 3SAT} and {\bf TSP} to
describe how these two methods differ in their approach to
solving NP problems.
\end{abstract}

\section{Introduction} \label{sec1}

The mathematical study of the selection, grouping and
permutation of a (usually) finite set of objects is called
\emph{combinatorial optimization}.  Although real-world
combinatorial problems have been known for centuries, it was
not until the last 30 years---i.e., until computers because
readily accessible---that finding a solution actually became
practical.  For example, the traveling salesman problem (TSP)
requires one to specify the order in which a salesman should
visit a fixed set of cities.  The objective is to find that
order which minimizes the total distance traveled.  If $N$
cites are to be visited, a brute force search must evaluate
$N!$ possible tours.  Clearly, this is not practical for $N$
much greater than eight or so.

The real challenge of combinatorial optimization is to create
algorithms and techniques that can solve realistically sized
problems within a reasonable number of computational
time~\cite{lawler76}.  Most algorithms formulate a
combinatorial problem as a search problem.
Implicit is the idea that the solutions to  combinatorial 
problems reside in an abstract solution space and two solutions are 
neighbors if they differ by a single mutation of a problem 
parameter.  Associated with each solution is a real number that 
reflects fitness or quality of that solution.  This space and the 
associated fitness values form a fitness landscape\footnote{In 
practice, fitness will be  with respect to one or more attributes 
such as cost or power consumption; high fitness is associated with 
good values of the attribute.}.  Any algorithm that ``solves'' an 
combinatorial problem is therefore a search algorithm that explores 
the fitness landscape. 

Unfortunately, many real-world combinatorial problems
require such huge computational resources that
brute force search methods are useless; they simply take too much
time to find the optimal answer.  This has led researchers to
use search heuristics that yield an acceptable compromise: a
possibly lower quality answer but with a minimal search
effort.
Evolutionary Computation (EC) techniques
are at the forefront of this work and impressive results have
been achieved.  Nevertheless, these EC techniques still run on
classical computers that use the Von Neumann model.  
But recently an entirely new approach has surfaced with
potentially enormous consequences.  This new approach is
called \emph{quantum computing}, which relies on the
principles of quantum mechanics to evolve solutions.
Existing programs, written in conventional high-level
languages cannot run on these machines.  In fact, no one even
knows how to build one!  Nevertheless, several system
architectures have been proposed and algorithms are being
developed, albeit in abstract form.

It is interesting to compare how a quantum search, running on
a quantum computer, differs from an evolutionary search, 
running on a classical computer.  
However, the whole point of this comparison is 
\emph{not} to advocate one method over the other---its purpose
is to highlight the radically different philosophical
approaches.  (Besides, because no one has ever built a quantum
computer, there is no way any direct comparison can be made at
this time.  It is up to the reader to decide which approach
holds the most promise.)  
If nothing else, the reader should
come away with an appreciation for the total re-orientation in
thinking that quantum search will require.   

The paper is organized as follows.  
Section \ref{sec2} provides a broad overview of
evolutionary algorithms and quantum computing.  Because the
focus is on NP-complete and NP-hard problems, a formal
definition of these problem classes is also provided.  Section
\ref{sec3} compares two quantum search methods against
evolutionary algorithms for two well known NP-complete problems
and one NP-hard problem.   
Finally,
Section \ref{sec4} comments on the future of quantum
computing.

 
\section{Background} \label{sec2}

This section reviews evolutionary algorithms, quantum
computing, and algorithms complexity.  Each topic is
wholely contained in a separate subsection so the reader
may skip familiar material.

\subsection{Evolutionary Algorithms}

 This section gives a brief introduction to evolutionary algorithms
(EAs).  For brevity, the emphasis is on those characteristics
of specific interest to our work.  More general information on
EAs can be found elsewhere (e.g., see \cite{back97}).

\begin{figure}[htbp]
\centerline{\epsffile{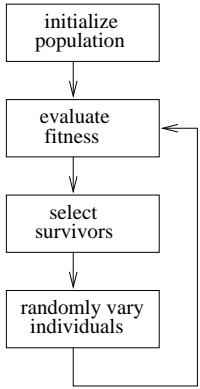}}
\caption{The canonical EA.  Each individual in the population
represents a unique solution to the optimization problem. 
The ``select survivors'' block
chooses highly fit parents for reproduction while the "randomly vary
individuals'' block generates the offspring (i.e., new
solutions for evaluation.) Selection may be deterministic or 
stochastic but 
generation of offspring is strictly stochastic. The process
continues until either a fixed number of generations have been
produced, or an acceptable solution has been found.}
\label{f1}
\end{figure}

Historically, evolutionary computations have a rich past, being
independently developed by at least three independent research 
efforts, which ultimately produced three distinct paradigms:
\emph{genetic algorithms}, \emph{evolutionary programming} and 
\emph{evolution strategies}.
All EAs share the same basic organization: iterations of competitive 
selection and random variation. More specifically, each generation 
(iteration of the EA) takes a population of individuals (potential 
solutions) and modifies the genetic material (problem parameters) 
to produce new individuals via stochastic operations.  
Both the parents and offspring are 
evaluated but only the highest fit individuals (better solutions)
survive over multiple generations.  Although there are 
several varieties of EAs, they are all biologically inspired 
and generally follow the format depicted in Figure \ref{f1}.

The evaluation function for an individual returns a numeric value 
representing the quality of the solution described by that
individual.  This numeric value is often called the \emph{fitness} 
of the individual while the evaluation function is called the 
\emph{fitness function}.  High fitness means the associated 
individual represents a good solution to the given problem.
The selection process targets highly fit individuals for
survival.

\subsection{Quantum Computing}

Almost twenty years ago Richard Feynman observed that
classical computers could not simulate certain quantum
mechanical effects \cite{feynman82}.  This observation
spawned interest in the field of quantum computing---i.e.,
computational machines that perform calculations by emulating
quantum mechanic effects.  Although no practical quantum
computer has yet been built (and the likelihood of building
one in the near future is bleak), the interest in this field 
has not diminished because of the enormous computational potential 
such a machine can provide.  Indeed, interest in this emerging
field is growing by leaps and bounds \cite{nsf}.

This section reviews some of the basic concepts of quantum mechanics
that relate to quantum computing.  No attempt has been made to be 
comprehensive and the reader is encouraged to study some of the 
excellent on-line tutorials~\cite{rieffel00,steane98,vedral98} or
journal reviews~\cite{spiller96}.  (However, see the
commentary on this topic in Section \ref{sec4}.)

Classical computer systems represent a single bit of information 
deterministically:  the value is either a logic 0 or a logic 1.
Quantum computer systems represent a single bit of information as a
\emph{qubit}, which is a unit vector in a complex Hilbert space $C^2$.  
The ideas are commonly
expressed using the bra/ket notation introduced by Dirac~\cite{dirac58}.
The \emph{ket} symbol is denoted by $\ketX$ and the
corresponding \emph{bra} is denoted by $\braX$.  The ket
describes a quantum state and the corresponding bra is its
complex conjugate.  

In computer science domains the ket (bra)
can be thought of as a column (row) vector.  That is, the orthonormal
basis $\{|0\rangle , |1\rangle\}$
can be expressed as $\{(0,1)^{\mbox{T}},
(1,0)^{\mbox{T}}\}$.  Any complex linear combination of two kets is also a 
ket.  The inner product of two vectors is denoted by $\langle x|y\rangle$.
Note that since $|0\rangle$ and $|1\rangle$ are orthonormal, $\langle 0|1\rangle= 0$.
$|x\rangle\langle y|$ denotes the outer product of the vectors. 

Any practical quantum computer must manipulate
a register of $n$ qubits.   If each qubit has an orthonormal
basis $\{|0\rangle,|1\rangle\}$, then a $n$ qubit system has a basis expressed by 
the \emph{tensor product}: $C^2\otimes C^2\otimes \cdots C^2$.  This gives $2^n$ basis 
vectors of the form 
\[
\ketzero\otimes\ketzero\otimes\cdots\ketzero
\]
\[
\ketzero\otimes\ketzero\otimes\cdots\ketone 
\]
\[ \vdots \]
\[
\ketone\otimes\ketone\otimes\cdots\ketone 
\]
In general, $|a\rangle$ denotes the tensor product 
$|a_n\rangle\otimes|a_{n-1}\rangle\otimes\cdots\otimes |a_1\rangle\otimes
|a_0\rangle$, which means a quantum register has the value $a=2^0a_0+2^1a_1+\cdots 2^na_n$.

A qubit need not exist in only one basis state.  Indeed, a qubit can exist as a
\emph{linear superposition} of basis states $c_0\ketzero + c_1\ketone$, where $c_0, c_1$ are 
complex numbers with $|c_0|^2+|c_1|^2=1$.  More generally, the $n$ qubit register can be 
prepared in a superposition of all possible classical states:
\begin{equation}
|x\rangle \; = \; \sum_{i=0}^{2^n-1} c_i|i\rangle
\label{eq1} 
\end{equation}
where the normalization condition 
\begin{equation}
\sum_i c_i^2 = 1
\label{eq2}
\end{equation}
must hold.  The complex number $c_i$ is called the \emph{amplitude} associated with the
state $|i\rangle$.

The linear superposition of states is key to understanding how quantum computers operate.  This
linearity feature means that any operation on a superposition of states renders the superposition of 
that operation on each state individually~\cite{hogg96}.  
There is no analogue in classical computer system for this principle and, as will be shown 
below, it is an important ingredient of the power behind quantum computing.   However, superposition
also permits the following rather bizarre situation.  Consider the state 
\[
\frac{1}{\sqrt{2}}(|00\rangle+|11\rangle)
\]
\emph{This state cannot be expressed in terms of the individual qubit states.}  The proof is
straightforward.  Note that 
\[
\begin{array}{ccc}
(a_1\ketzero+b_1\ketone)\otimes(a_2\ketzero+b_2\ketone) & = &  
a_1a_2|00\rangle + a_1b_2|01\rangle + b_1a_2|10\rangle + b_1b_2|11\rangle \\
& = & |00\rangle + |11\rangle
\end{array}
\]
Clearly $a_1b_2 = 0$, but this implies either $a_1a_2$ or $b_1b_2$ must equal zero,
which is not possible.  States that cannot be described by individual qubit states
are called \emph{entangled}.  There is considerable debate concerning the actual role entanglement
plays in search operations.  This issue will be discussed in more depth in Section \ref{sec3}.

The state of a qubit register is determined by a measurement.  In quantum systems this measurement
process projects the system state onto one of the basis states.  Referring to Eq.~(\ref{eq1}),
the measurement returns a value of $|i\rangle$ with probability $|c_i|^2$.  Any subsequent measurement
returns the state $|i\rangle$ with probability 1, which means the measurement process irreversibly 
alters the state of the system.  Measurement also gives another perspective on entanglement: two 
qubits are entangled if and only if the measurement of one effects the state of the other.

A quantum computer can perform the same
function $f$ as a classical computer if that function is a one-to-one mapping from the 
domain to the range.  In other words, $f$ must be a reversible function.  Reversibility is usually
mentioned in the context of performing computations without expending heat~\cite{bennett73}.  Here, however, 
reversibility must hold or $f$ will not be physically realizable on a quantum computer. Hogg~\cite{hogg96}
illustrates the importance of reversibility with a simple example.  Suppose $f(s_1)= f(s_2) = s_3$.  
Then for the superposition $|s\rangle = \frac{1}{\sqrt(2)}(|s_1\rangle + |s_2\rangle)$ linearity forces
$f(|s\rangle )= \frac{1}{\sqrt(2)}(|f(s_1)\rangle + |f(s_2\rangle)$.  But this equals $\sqrt(2)|s_3\rangle$,
which violates the normalization condition given in Eq.~(\ref{eq2}).  

Quantum systems evolve from state to state according to Schr\"{o}dinger's equation~\cite{feynman65}.  
Vector states can be expressed as a superposition of basis states each having an amplitude $|\psi_i\rangle$.
This means evolution occurs by modification of the state amplitudes.
Clearly, we would like to increase the amplitude of that state with the desired answer.  Suppose we start in state 
$|a\rangle=\sum \psi_k |a_k\rangle$.  This system evolves over time under a linear operator 
$U$, i.e., $|a'\rangle = U|a\rangle = \sum \psi_k' |a_k\rangle$.   Hence, $\psi' = U\psi$ and Eq.~(\ref{eq2})
is satisfied iff $U$ is unitary.  To see this,  consider the inner product $(\psi')^{\dagger}\psi'$, which 
must equal one because state
vectors are orthonormal.  
Substituting $\psi' = U\psi$ yields
\[
(U\psi)^{\dagger}(U\psi) \; = \; \psi^{\dagger}(U^{\dagger}U)\psi
\]
This inner product equals one if $(U^{\dagger}U)= I$.  Hence, $U$ must be unitary.

It is convenient to adopt a simplified notation when describing unitary operations that
are applied to individual qubits.  Some common unitary operators are
\begin{eqnarray*}
I: &  \ketzero\rightarrow \ketzero \\
& \ketone\rightarrow\ketone 
\end{eqnarray*}
\begin{eqnarray*}
X: &  \ketzero\rightarrow \ketone \\
& \ketone\rightarrow\ketzero 
\end{eqnarray*}
\begin{eqnarray*}
Z: &  \ketzero\rightarrow \ketzero \\
& \ketone\rightarrow -\ketone 
\end{eqnarray*}
where $I$ is an identity operator, $X$ a negation operator, and $Z$ a phase shift
operator. Suppose we have a 3 qubit register and we want to negate the first qubit
and leave the other qubits unaltered.  This transformation is denoted by 
$X\otimes I\otimes I$.  

An extremely important transformation is the \emph{Walsh-Hadamard transformation} defined
as
\begin{eqnarray*}
H: &  \ketzero\rightarrow \frac{1}{\sqrt{2}}(\ketzero + \ketone) \\
& \ketone\rightarrow \frac{1}{\sqrt{2}}(\ketzero - \ketone)   
\end{eqnarray*}
When applied to $\ketzero$,
a superposition state is created.  When applied to $n$ bits individually, a superposition
of all $2^n$ states is created.  Specifically,
\[
(H\otimes H \otimes\cdots\otimes H)|000\cdots 0\rangle
\]
\[
= \frac{1}{\sqrt{2^n}}((\ketone + \ketzero)\otimes(\ketone + \ketzero)\otimes\cdots\otimes(\ketone + \ketzero))
\]
\[
= \frac{1}{\sqrt{2^n}}\sum_{x=0}^{2^n-1} |x\rangle
\]

It is important to emphasize the role superposition plays in quantum computing.  Let $U_f$ be a unitary
transformation corresponding to a classical function $f$, i.e., $U_f: |x\rangle|y\rangle\rightarrow
|x\rangle|y\oplus f(x)\rangle$, where $\oplus$ represents bitwise exclusive-or.  Notice that this 
transformation preserves the input---which must be done if $f$ is not invertible---thereby
making $U_f$ unitary~\cite{deutsch85}.  We can
think of $|y\rangle$ as the hardware of the quantum computer.
When this $U_f$ operates on a superposition of states as in Eq.~(\ref{eq1}), the result is
\begin{eqnarray*}
U_f(\sum_{i=0}^{2^n-1}c_i|i\rangle|0\rangle) & = & \sum_{i=0}^{2^n-1} c_iU_f(|i\rangle|0\rangle) \\ 
& = & \sum_{i=0}^{2^n-1} c_i|i\rangle|0\oplus f(i)\rangle \\
& = & \sum_{i=0}^{2^n-1} c_i|i\rangle|f(i)\rangle 
\label{eq3} 
\end{eqnarray*}
Notice that $f$ is simultaneously applied to all basis vectors.  \emph{Hence, a single application of $U_f$
computes all $2^n$ values of $f(0),\ldots,f(2^n-1)$ at once}~\cite{barenco98}.  It is this quantum parallelism
that is primarily responsible for the enormous interest in quantum computing.  But something must be
wrong.  How can you extract an exponential amount of information out of a linear number of qubits?  The 
answer lies with the amplitudes.  If $c_i= c_j \; \forall i,j$, then a measurement will produce any of the $2^n$ states 
with equal probability.  Furthermore, once that measurement is taken, the system collapses into that measured state and
all other information is lost.  (You really can't get something for nothing.)  Nevertheless, you can exploit this 
parallelism using the property of \emph{quantum interference}.  Interference allows the exponential number of computations 
performed in parallel to either cancel or enhance each other.   Feynman~\cite{feynman65} beautifily describes how light 
waves can constructively or destructively interfere to produce
this effect.  The goal of any quantum algorithm is to have a similar phenomena occur---i.e., interference increases the amplitudes of computational results we desire and decreases the amplitudes 
of the remaining results.  It is a unitary operator that would alter these amplitudes.  Examples of this approach are
presented in Section \ref{sec3}.

\subsection{NP-Complete vs.~NP-Hard Problems} 

Many papers that discuss NP-complete and NP-hard problems (incorrectly) 
presume the 
reader fully understands the difference between these two labels.
The distinction is important.
For example, if one formulates a {\bf TSP} problem as ``does a
tour exist of length $\le k$'', then this is NP-complete.
However, if the problem asks ``what is the minimum length
tour'', then this problem is NP-hard because it isn't in class
NP.  (They answer is not verifiable in polynomial time.  The only
way of answering `yes' is to enumerate all possible tours.)
In this paper a more formal approach is taken: all terms and three
example problems are formally defined.
This material is primarily taken from \cite{manber89}.
I begin with the following basic definitions:

\begin{defn}  (decision problem) \\
A problem in which the only answer is either YES or NO.
\end{defn}

\begin{defn} (language) \\
The set of all possible input strings to a decision problem that render a YES answer.
\end{defn}

The input strings are defined some fixed alphabet of symbols.  For example, binary
strings are defined over the alphabet $\{0,1\}$.

\begin{defn} (polynomial-time algorithm) \\
An algorithm that completes execution in a time which is a 
polynomial factor
of the size of its input parameters.
\end{defn}

\begin{defn}
(polynomially reducible) \\
Let $L_1$ and $L_2$ be two languages.  $L_1$ is polynomially reducible to $L_2$ (denoted
by $L_1\propto L_2$) if there exists some polynomial-algorithm that converts every input
instance $i_1\in L_1$ into another $i_2\in L_2$.
\end{defn}

It should be stressed that reducibility is asymmetric.  In
other words, if $L_1\propto L_2$, then this does not necessarily
mean $L_2\propto L_1$.  Nevertheless, polynomial reducibility
does have an important characteristic, which is given in the
following theorem:
\begin{guess}
If $L_1\propto L_2$, and there is a polynomial-time algorithm for $L_2$, there there
is a polynomial-time algorithm for $L_1$. (See~\cite{manber89}, page 343 for proof.)
\end{guess}

\begin{defn}
(nondeterministic algorithm) \\
An algorithm that permits more than one possible move at some step during its execution.
\end{defn}

With these definitions it is now possible to define the algorithm classes P and NP.

\begin{defn} (class P) \\
P is the class of languages (decision problems) $L$ that, with input $x$, can
 in polynomial time return an answer YES if and only if $x\in L$. 
\end{defn}

\begin{defn} (class NP) \\
NP is the class of languages (decision problems) that can be checked for correctness
in polynomial time.
\end{defn}

 Notice that the above definition says
nothing about the computational effort required to get that answer---it merely says
to verify the correctness of an answer takes only polynomial time.  
Whether or not P=NP has yet to be determined.

It is now possible to formally define NP-hard and NP-complete.  It should be
emphasized that
the two type of problem classes are \emph{not} interchangeable.

\begin{defn}
(NP-hard) \\
A problem $\cal P$ is NP-complete if  
every other problem in NP is polynomially reducible to $\cal P$
\end{defn}

\begin{defn} (NP-complete) \\
A problem $\cal P$ is NP-complete if (1) $\cal P\in$NP, and (2)
every other problem in NP is polynomially reducible to $\cal P$
\end{defn}

NP-complete problems are decision problems.  NP-hard problems ask 
for the optimal solution to an NP-complete problem.  And, they have 
at least the same level of difficulty to solve as does the 
corresponding NP-complete problem.  There are a very large number of 
problems that have been proven to be 
NP-complete\footnotemark.  Theorem 2 shows the most common way of 
proving a decision problem is NP-complete:
\begin{guess}
A problem X is an NP-complete problem if (1) X belongs to NP, and (2) Y
is polynomially reducible to X, for some problem Y that is NP-complete.
(See \cite{manber89}, page 346 for proof).
\end{guess}

Theorem 2 is also used to prove if a problem $X$ is
NP-hard.  Consider a problem $X=(D,\eta)$ that has an input
domain $D$ and some property $\eta$.  An algorithm which
solves problem $X$ uses an input instance $I\in D$ and
verifies whether $\eta$ holds for input $I$.  Suppose there
exists another problem $X' = (D', \eta')$ for which $X\propto X'$.
This means (1) $I\in D$ can be transformed to $I'\in D'$ in
polynomial time, and (2) for any $I\in D$, $\eta$ holds if and
only if $\eta'$ (based on $I'\in D'$) holds.  $X'$ will be
NP-hard if $X$ is NP-complete and $X\propto X'$.

I now describe two known NP-complete problems~\cite{garey79}, which
will be used in Section \ref{sec3}. 
Let $\Sigma$ be a Boolean expression in Conjunctive Normal Form (CNF)---i.e.,
$\Sigma$ is the logical \emph{and} of two or more clauses where each clause 
is the logical \emph{or} of Boolean variables or their complements.  An example
is $\Sigma = (x + y + \overline{z})\cdot(\overline{x} + \overline{y})\cdot(\overline{y} + z)$.
This Boolean expression is considered \emph{satisfied} if an assignment of
0s and 1s to the Boolean variables makes $\Sigma$ equal to a logic 1.

\vspace{0.1in}

{\textsf{\bf SATISFIABILITY PROBLEM (SAT)}:}

\emph{Instance:} a Boolean expression in CNF

\emph{Question:} Does there exist an assignment of 0s and 1s to 
the variables such that the expression is satisfied? 
 
\vspace{0.1in}

{\textsf{\bf 3SAT PROBLEM}:}

\emph{Instance:} a Boolean expression in CNF with each clause having 
exactly three variables

\emph{Question:} Does there exist an assignment of 0s and 1s to 
the variables such that the expression is satisfied? 

\vspace{0.1in}

Finally, I describe a famous problem, which will
also be used in Section \ref{sec3}.  This problem is
NP-hard~\cite{garey79}:

\vspace{4ex}

{\textsf{\bf TRAVELING SALESMAN PROBLEM (TSP)}:}

\emph{Instance}: a finite set $C=\{c_1, c_2,\ldots, c_m\}$ of cities, and a distance 
$d(c_k , c_j)\in Z^+$ for each pair of cities $c_k , c_j\in C$.  
 
\emph{Question}:  What permutation $[c_{\pi(1)}, c_{\pi(2)},\ldots, c_{\pi(m)}] $ of $C$ will minimize the tour length

\begin{displaymath}
\left\{\sum_{i=1}^{m-1} d(c_{\pi(i)}, c_{\pi(i+1)})\right\} + d(c_{\pi(m)}, c_{\pi(1)}) \; ?
\end{displaymath}

\footnotetext{A large database can be found at
http://www.nada.kth.se/$\tilde{\hspace{1pt}}$viggo/problemlist/compendium.html.}

\section{Search Approaches} \label{sec3}

This section provides examples of how evolutionary and
quantum search approaches have been applied to NP problems.
No attempt has been made to survey the field; the objective
is to present a few examples so the reader can appreciate
the philosophical differences, which will be discussed in depth
in  Section \ref{sec4}.

\subsection{Evolutionary Search}

Although a number of papers have appeared discussing attacking 
{\bf SAT} problems using EAs, I will focus on the recent work
by B\"{a}ck, et al~\cite{back98}.  They used an evolution strategy
to find solutions to instances of the {\bf 3SAT} problem.

The search for a satisfiable solution is difficult because, as the
authors point out, the fitness landscape is extremely flat---any genetic
search reverts to a random search.  Moreover, this type of landscape
makes it difficult to define fitness in a meaningful way.  The authors
get around this situation by adapting a method suggested by a colleague~\cite{mich}.
This alternative method replaces each literal with $x$ with $(y-1)^2$ and $\overline{x}$
with $(y+1)^2$.  Furthermore, each disjunction $\wedge$ is replaced by an arithmetic + (sum)
and each conjunction $\vee$ is replaced by an arithmetic  $\cdot$ (product).  The 
resulting fitness function has a minimum of 0, when the $y_i$'s converge to 1 (true) or
-1 (false).  These changes convert {\bf 3SAT} into a real-parameter optimization problem,
which evolution strategies are ideally suited for.

The evolution strategy randomly initialized the object parameters to values between
-1.0 and 1.0.  A (15,100)-ES with one standard deviation ($\sigma$) was used; $\sigma$
had an upper limit of 3.0.  Later versions introduced various forms of recombination, 
which ultimately was shown to render the best performing version.

\subsection{Quantum Search}

Quantum search approaches differentiate between \emph{structured}
problems, where partial solutions can be extended to complete
solutions, and \emph{unstructured} problems.  The unstructured
approach can be used for finding solutions for NP-hard problems.

\subsubsection{NP-Complete Problems}

Ohya and Masuda~\cite{ohya98} developed a quantum search method 
that is frequently used for NP-complete problems. 
Their algorithm starts with the quantum system in the state 

\[
|s\rangle \; = \; \frac{1}{\sqrt{2^n}}\sum_{x_1,\ldots,x_n=0}^1 \otimes_{j=1}^n|x_j\rangle \otimes_1^k\ketzero\otimes\ketzero
\]
for a {\bf SAT} instance with variables $x_1,\ldots,x_n$.  The $k$ qubits are garbage
bits needed by reversible logic gates and the final qubit (initialized to $\ketzero$) 
indicates if the expression is satisfied.  Then, using a unitary operator $U_f$,  
\begin{eqnarray*}
|t\rangle & = & U_f|s\rangle \\
 & = &  \frac{1}{\sqrt{2^n}}\sum_{x_1,\ldots,x_n=0}^1 U_f\otimes_{j=1}^n|x_j\rangle 
\otimes_1^k\ketzero\otimes\ketzero \\
 & = &  \frac{1}{\sqrt{2^n}}\sum_{x_1,\ldots,x_n=0}^1 \otimes_{j=1}^n|x_j\rangle 
\otimes_{m=1}^k|y_m\rangle\otimes|f(x_1,\ldots,x_n)\rangle \\
\end{eqnarray*}
where $f()$ is the Boolean expresssion.
The last qubit is then measured by applying a projector $P= I\otimes|1\rangle\langle 1|$
to $|t\rangle$.   If out of the $2^n$ possible solutions there are $r$ solutions that 
satisfy $f$, then the probability of measuring a solution is 
$|P|t\rangle|^2 = r/2^n$.  For small $r$ this probability is quite small.  Hence, in
practice quantum search algorithms try to exploit quantum interference to amplify the 
amplitude of the desirable solutions and attenuate all other amplitudes.

Cerf, et al.~\cite{cerf98} provide a good description of exactly how this is done.
Their approach
relies on an ``oracle'' function $f(x)$ that equals one for the optimal
input $x$ (and zero elsewhere).  The goal is for the quantum system to evolve
from an initial state $|s\rangle$ to the target state $|t\rangle$ in minimum
time.  Note that $f(x)=1$ only at $x=t$.  More precisely, the goal is to
increase the amplitude of $|t\rangle$ to a point where a measurement
will render $|t\rangle$ with the highest probability.

Assume an arbitrary unitary operator $U$ has been found that 
connects $|s\rangle$ to $|t\rangle$---i.e., $\langle t|U|s\rangle \neq 
0$.  The probability $|t\rangle$ is actually found 
is $|\langle t|U|s\rangle|^2$, which means the experiment must be
repeated $|\langle t|U|s\rangle|^{-2}$ times on average to guarantee 
success.  However, it is possible to reduce this to the order 
of $|\langle t|U|s\rangle|^{-1}$---which can
be a considerable savings---with an appropriate quantum search algorithm.

The algorithm begins in a superposition of states and any measurement is postponed
until the end.  Cerf, et al.~\cite{cerf98} defined a specific unitary operator 
\[
Q = -Ue^{i\pi P_s}U^{\dagger}e^{i\pi P_t}
\]
where $P_s = |s\rangle\langle s|$ and $P_t = |t\rangle\langle t|$ are projection
operators on $|s\rangle$ and $|t\rangle$.  These exponential operators simple 
flip the phase on a state.  For example, the phase of state $|x\rangle$ is flipped
by $e^{i\pi P_s}$ iff $x=s$.  Since the objective is search for state $|t\rangle$,
the oracle is used to implement its exponential operator.  That is, 
$e^{i\pi P_t}|x\rangle = (-1)^{f(x)}|x\rangle$.  Then, by repeatedly applying $Q$,
the amplitude of $|t\rangle$ is amplified, beginning at $U|s\rangle$.  
This amplitude amplification is achieved by the repeated application of $Q$ which,
in effect, rotates the starting state $|s\rangle$ into the target state $|t\rangle$.
In other words, the beginning state $U|s\rangle$ is rotated to the target state $|t\rangle$
by repeated applications of $Q$, followed by a measurement.  Recall $U$ was an 
\emph{arbitrary} unitary operator; by using structure information it may be possible
to find a better $U'$ so that $U'|s\rangle$ has larger amplitudes in states which are
more probable to be solutions.  Cerf, et al.~\cite{cerf98}
describe a method that constructs such a $U'$.
 
Grover's quantum search algorithm searches a random database
of $N$ items in $O(\sqrt(N))$ steps~\cite{grover97}.  This means 
\underline{unstructured} NP-complete problems can be solved by forming a database
of all possible candidate solutions, and then use Grover's algorithm
to find the solution.  Although this is a considerable speedup over
classical machines, it may not be all that impressive.   For instance, if one 
has to find an assignment of one of $k$ values 
to $n$ total variables, a classical algorithm would take $O(k^n)$ steps while  
quantum algorithms would still take $O(k^{n/2})$ steps.  Nevertheless, the algorithm
does find a use with both NP-complete and NP-hard problems.

\subsection{NP-Hard Problems}

\begin{figure}[htbp]
\centerline{\epsffile{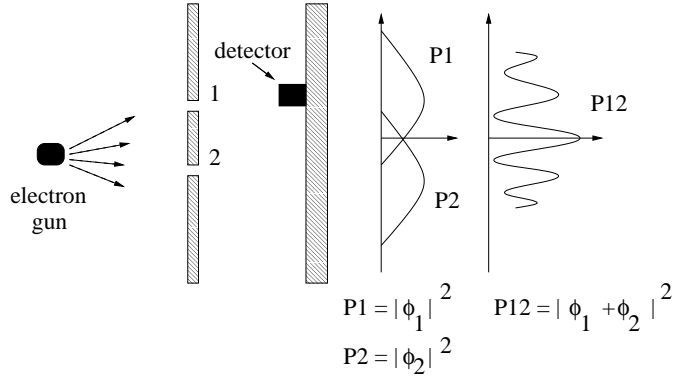}}
\caption{An interference experiment showing how amplitudes combine
in both a constructive and destructive manner.  The electron gun fires
electrons that go through slits in a wall.  A
movable detector determines where the electrons impact the backstop.
The wavelike behavior of the electrons produces interference so that
the total distribution $P_{12}\ne P_1 + P_2$.  This figure was adopted
from \cite{feynman65}.}
\label{slit}
\end{figure}

A beautiful example of how non-traditional architectures can solve NP-hard
problems is the scheme presented by \v{C}ern\'y~\cite{cerny93} to solve an
instance of {\bf TSP}.  Figure \ref{slit}
shows an interference experiment setup.

\begin{figure}[htbp]
\centerline{\epsffile{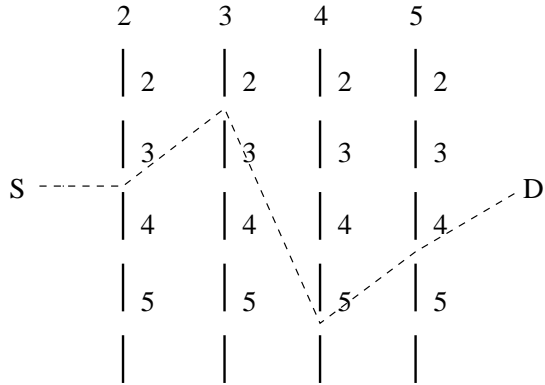}}
\caption{A sample TSP trajectory in the \v{C}ern\'y quantum
computer. This tour is $S,3,2,5,4,D$.  Note that not all tours
are ``legal''.  For instance, $S,3,3,5,4,D$ is also a trajectory
but it is illegal because city 3 is visited twice and city 2
is never visited.}
\label{traj}
\end{figure}

 \v{C}ern\'y proposed a quantum computer similar to that of the interference
experiment.  This computer has ($n$-1) walls representing cites 2,3,$\ldots,n$.
Furthermore, each wall has ($n$-1) slits.  A beam of quanta (e.g., electrons)
sent through
this array has ($n$-1)$^{n-1}$ possible trajectories.  The wavelike behavior
of electrons means a superposition of all possible trajectories is rendered
in $O(n)$ time.  A sample trajectory in this quantum computer
is shown in Figure \ref{traj}.

These trajectories identify tours but they do not indicate the
length of those tours.  Since this machine is hypothetical, an
internal degree of freedom can be added---even if
nature doesn't provides it.  Specifically,  the internal state
of a particle is 
\[
|k; c_2, c_3, \ldots, c_n; p\rangle
\]
where $k\in\{0,1,2,\ldots,NL\}$, $c_i\in\{0,1\}$, and
$p\in\{0,1\}$.  The quantum number $k$ measures the tour
length; $c_i=0$ if city $i$ is not visited and 1 otherwise;
and the quantum number $p$ is used to
control the dynamics of the search~\footnotemark.

\footnotetext{See the appendix in~\cite{cerny93} for an explanation 
of how $p$ is used; its purpose is not needed for the brief overview 
given in this paper.}

To illustrate the dynamics, let
$(i,m)\rightarrow(i+1,n)$ denote a trajectory between slots on
neighboring layers $i$ and $i+1$ indicating the tour moves
from city $m$ to city $n$.  If the particle moves through slot
$(i,n)$, then 
\[
c_n=0 \rightarrow c_n=1.
\]
Furthermore, assume the particle moving between layers
encounters a field that increases the quantum number $k$ by
a factor $d_{mn}$ if the trajectory moves from ($i,m)$ to 
$(i+1,n)$, where $d_{mn}$ is the distance between the two cites.
Then, with an initial state
\[
|0;0,0,\ldots,0;0\rangle
\],
after passing through the machine the particles are in a state
\[
\sum_{\mbox{trajectories}} 
|k; c_2, c_3, \ldots, c_n; p\rangle_{\mbox{trajectory}}.
\]
Note that the legal tours will have all $c_i=1$ and the
quantum number $k$ is the tour length.  A filter installed at
point $D$ purges all kets with at least one $c_i=0$.  This
leaves   
\begin{equation}
\sum_{\mbox{TS routes}} 
|k; c_2, c_3, \ldots, c_n; p\rangle_{\mbox{TS route}}.
\label{eqtsp}
\end{equation}
A Stern-Gerlach-like device~\cite{feynman65} could be used to
construct such a filter, which would
split the above superposition into $NL$ streams according to
the $k$ value.  A set of particle detectors would then
indicate the tour length---i.e., a detector measuring $M$
would fire if there exists a TS tour with a length equal to
$M$.

It is important to emphasize this does \emph{not} mean an 
instance of {\bf TSP} can be solved in polynomial time.  In
principle, Eq.~(\ref{eqtsp}) could represent a superposition of
$O(n!)$ states.  Hence, even if Grover's algorithm is used, 
it would take $O(\sqrt{n!})$ steps to find the minimal length tour.


\section{Discussion} \label{sec4}

No one has yet built a quantum computer capable of searching
for solutions to even moderate size NP problems.  
But, despite our inability to make head-to-head comparisons of
evolutionary and quantum searches, it is possible to highlight
their primary philosophical difference:

\begin{description}
\item{\emph{evolutionary search}}
 
The algorithm uses stochastic operations to explore a fitness
landscape comprised of all possible solutions.

\item{\emph{quantum search}}

The algorithm forms a superposition of all possible problem states
and then applies a unitary operator to compute a superposition of 
all possible solutions.  This unitary operator also 
alters the amplitudes of each state exploiting interference
to maximize the amplitudes of the desired states.  A final 
measurement extracts the solution with a probability equal to
the amplitude squared. 

\end{description}

EAs must tradeoff \emph{exploration} against \emph{exploitation}.
In other words, the EA must carefully decide which regions of the 
fitness landscape to abandon, because the solutions
are found there are poor, without putting much emphasis on regions 
with good solutions because that would limit the search.  The focus 
of EA research with respect to NP problems is in two areas: (1) 
identification of appropriate representations of the problem 
parameters, which ultimately defines the 
fitness landscape, and (2) creation of effective stochastic
reproduction operators that control movement over the fitness landscape.

Quantum search algorithm exploit superposition to produce massive 
parallelism.  One rather contentious debate in this field is the role 
entanglement plays.  On one side of the fence are those who feel 
entanglement is essential for speedup~\cite{braun00}, while on the 
other side are those who feel it is completely 
unnecessary~\cite{lloyd99a,knight00}.  The latter group believes
superposition and interference are sufficient to produce speedup.  
This issue could be resolved if a truly entangled system were available
for study.  Unfortunately, recent room-temperature liquid-state
NMR experiments have failed to produce any entangled states.  Still, 
some researchers feel increasing the number of qubits (currently only 
around 3) will eventually make entanglement appear~\cite{fitz00}.  

One of the main difficulties in running a quantum computer is they
must remain completely isolated from their environment or 
the state evolution will cease.  Furthermore, there is no way
of observing what's going on unless a measurement is taken.
But taking a measurement process changes the system by causing
it to collapse into
one of the basis states.  Some methods of dealing with this
have been proposed~\cite{feynman82}, but it still remains a
thorny issue.  Consequently, we can expect running a quantum
search will be much more fragile than the running of an 
evolutionary search on a classical computer.  
 
One final note on unstructured NP-complete problems. 
The $O(\sqrt(N))$ time for Grover's search algorithm has been proven 
to be optimal~\cite{zalka97}.  This has a rather disappointing 
consequence: if $O(\sqrt(N))$ time is optimal, this may mean quantum 
computers can't solve NP problems with an exponential speedup.
Preskill~\cite{preskill97} suggests the 
real application area may lay outside NP.  Quantum system simulation 
is one example, which was also previously suggested by 
Feynman~\cite{feynman82}.
 

\section{Final Comments} \label{sec5}

I will conclude this paper with some personal observations.
I do believe quantum computing will change the way computer engineers and scientists
think about computing systems.  To date, quantum computing has been the
domain of primarily physicists.  It is about time that computer engineers
and scientists enter this arena and begin to drive its direction.

Many computer professionals entering this field are quickly put off by
the notational and conceptual barriers.  Tutorials are available (e.g.,
see \cite{vedral98, aharonov98, lomonaco00, rieffel00}), but many 
readers will quickly find them incomprehensible---they are written by 
physicists for
physicists.  (Out of this lot, however, I believe \cite{rieffel00} is the
best.)  The sad truth is a computer professional who lacks a firm foundation---i.e., 
formal training---in quantum theory will most likely not be
able to contribute to the quantum computing field.  As an absolute minimum
I would recommend an upper division undergraduate
course in quantum mechanics.  This should
be sufficient background for one to begin reading the literature from the field.

My other observation concerns the practicality of the currently proposed quantum
computer architectures.  Many proposed systems (including the NMR approaches)
contain a vast network of interconnected quantum gates (such
as AND gates) which implement some function 
$f(x)$~\cite{grover99}.  I believe this is entirely too low of a 
design level, which is unlikely to lead to massive
improvements in computation power---certainly no where near
orders of magnitude improvement.  
Although, in principle, all computer systems are just 
interconnected primitive logic
gates, engineers typically do not visualize them in this way.  For
example, few designers think of a processor as a
network of primitive logic gates implementing 
Boolean expressions.  For the most part architectural
design work is rarely performed
lower than the register-transfer level.  Indeed, the hardware description 
languages in use today, such
as VHDL and Verilog, are most frequently used at the register-transfer level. 

Thinking of quantum computers in terms of interconnected logic
gates also tends to limit their ability to perform general
purpose computations---especially those computations that are
inherently parallel.  For instance, can a quantum computer 
perform an evolutionary search?  

I am convinced that a radical increase in computing power will 
only come once the Von Neumann paradigm has been dispensed with.  
Architectures such as those proposed by \v{C}ern\'y~\cite{cerny93} 
are an example of the imagination that will be required.


\bibliography{qm}

\begin{thebibliography}{10}

\bibitem{lawler76}
E.~Lawler.
\newblock {\em Combinatorial Optimization: Networks and Matroids}.
\newblock Holt, Rinehart and Winston, 1976.

\bibitem{back97}
T.~B\"ack, U.~Hammel, and H.-P. Schwefel.
\newblock Evolutionary computation: comments on the history and current state.
\newblock {\em IEEE Trans. Evol. Comp.}, 1:3--17, 1997.

\bibitem{feynman82}
R.~Feynman.
\newblock Simulating physics with computers.
\newblock {\em Intl. J. Theo. Phys.}, 21:467--488, 1982.

\bibitem{nsf}
National~Science Foundation.
\newblock Quantum information science.
\newblock {\em Report of the NSF Workshop, Arlington, VA}, 1999.

\bibitem{rieffel00}
E.~Rieffel and W.~Polak.
\newblock {\em An introduction to quantum computing for non-physicists}.
\newblock Los Alamos Physics preprint archive,
  http://xxx.lanl.gov/abs/quant-ph/9809016, 2000.

\bibitem{steane98}
A.~Steane.
\newblock {\em Quantum Computing}.
\newblock Los Alamos Physics preprint archive,
  http://xxx.lanl.gov/abs/quant-ph/9708022, 1998.

\bibitem{vedral98}
V.~Vedral and M.~Plenio.
\newblock {\em Basics of quantum computation}.
\newblock Los Alamos Physics preprint archive,
  http://xxx.lanl.gov/abs/quant-ph/9802065, 1998.

\bibitem{spiller96}
T.~Spiller.
\newblock Quantum information processing: cryptography, computation, and
  teleportation.
\newblock {\em Proc. of IEEE}, 84:1719--1746, 1996.

\bibitem{dirac58}
P.~Dirac.
\newblock {\em The Principles of Quantum Mechanics}.
\newblock Oxford University Press, 4th edition, 1958.

\bibitem{hogg96}
T.~Hogg.
\newblock {\em Quantum computing and phase transitions in combinatorial
  search}.
\newblock Los Alamos Physics preprint archive,
  http://xxx.lanl.gov/abs/quant-ph/9508012, 1996.

\bibitem{bennett73}
C.~Bennett.
\newblock Logical reversibility of computation.
\newblock {\em IBM J. Res. Dev.}, 17:525--532, 1973.

\bibitem{feynman65}
R.~Feynman, R.~Leighton, and M.~Sands.
\newblock {\em Lectures on Physics, Vol. III}.
\newblock Addison-Wesley, 1965.

\bibitem{deutsch85}
D.~Deutsch.
\newblock Quantum theory, the \mbox{C}hurch-\mbox{T}uring principle and the
  universal quantum computer.
\newblock {\em Proc.~Royal Soc.~of London A}, \mbox{A}400:97--117, 1985.

\bibitem{barenco98}
A.~Barenco.
\newblock {\em Quantum computation: an introduction}.
\newblock in \emph{Introduction to Quantum Computation and Information}, H. Lo,
  S. Popescu and T. Spiller (Eds.), World Scientific, 1998.

\bibitem{manber89}
U.~Manber.
\newblock {\em Introduction to Algorithms}.
\newblock Addison-Wesley, 1989.

\bibitem{garey79}
M.~Garey and D.~Johnson.
\newblock {\em Computers and intractability: a guide to the theory of
  NP-completeness}.
\newblock W.H. Freeman \& Company, 1979.

\bibitem{back98}
T.~B\"{a}ck, A.~Eiben, and M.~Vink.
\newblock A superior evolutionary algorithm for \mbox{3SAT}.
\newblock {\em Proc.~\mbox{EP}98}, 1998.

\bibitem{mich}
Z.~Michalewicz.
\newblock personal communication with authors of \cite{back98}.

\bibitem{ohya98}
M.~Ohya and N.~Masuda.
\newblock {\em NP problem in quantum algorithm}.
\newblock Los Alamos Physics preprint archive,
  http://xxx.lanl.gov/abs/quant-ph/9809075, 1998.

\bibitem{cerf98}
N.~Cerf, L.~Grover, and C.~Williams.
\newblock {\em Nested quantum search and \mbox{NP}-complete problems}.
\newblock Los Alamos Physics preprint archive,
  http://xxx.lanl.gov/abs/quant-ph/9806078, 1998.

\bibitem{grover97}
L.~Grover.
\newblock Quantum mechanics helps in searching for a needle in a haystack.
\newblock {\em Phy. Rev. Ltr.}, 79:325--328, 1997.

\bibitem{cerny93}
V.~\v{C}ern\'y.
\newblock Quantum computers and intractable (\mbox{NP}-complete) computing
  problems.
\newblock {\em Phys.~Rev.~A}, 48:116--119, 1993.

\bibitem{braun00}
S.~Braunstein and A.~Pati.
\newblock {\em Speedup and entanglement in quantum searching}.
\newblock Los Alamos Physics preprint archive,
  http://xxx.lanl.gov/abs/quant-ph/0008018, 2000.

\bibitem{lloyd99a}
S.~Lloyd.
\newblock Quantum search without entanglement.
\newblock {\em Phys. Rev. A}, 61:R10301--01304, 1999.

\bibitem{knight00}
P.~Knight.
\newblock Quantum information processing without entanglement.
\newblock {\em Science}, 287:441--442, 2000.

\bibitem{fitz00}
R.~Fitzgerald.
\newblock What really gives a quantum computer its power?
\newblock {\em Physics Today}, pages 20--22, 2000.

\bibitem{zalka97}
C.~Zalka.
\newblock {\em Grover's quantum searching algorithm is optimal}.
\newblock Los Alamos Physics preprint archive,
  http://xxx.lanl.gov/abs/quant-ph/9711070, 1997.

\bibitem{preskill97}
J.~Preskill.
\newblock {\em Quantum computing: pro and con}.
\newblock Los Alamos Physics preprint archive,
  http://xxx.lanl.gov/abs/quant-ph/9705032, 1997.

\bibitem{aharonov98}
D.~Aharonov.
\newblock {\em Quantum computation}.
\newblock Los Alamos Physics preprint archive,
  http://xxx.lanl.gov/abs/quant-ph/9812037, 1998.

\bibitem{lomonaco00}
\mbox{J}r. S.~Lomonaco.
\newblock {\em A rosetta stone for quantum mechanics with an introduction to
  quantum computation}.
\newblock Los Alamos Physics preprint archive,
  http://xxx.lanl.gov/abs/quant-ph/0007045, 2000.

\bibitem{grover99}
L.~Grover.
\newblock Quantum mechanical searching.
\newblock {\em Proc.~\mbox{CEC}99}, pages 2255--2261, 1999.

\end{thebibliography}

\end{document}